\documentclass[useAMS,usenatbib]{mn2e}

\newcommand\arcdeg{\degr}%
\newcommand\lesssim{\la}%
\usepackage{graphicx}

\title[Brown Dwarf Search in Pleiades]
{An Optical and Near Infrared Search for Brown Dwarfs in the Pleiades Cluster}
\author[Chie Nagashima et al.]
{Chie Nagashima,$^{1}$\thanks{E-mail:chie@optik.mtk.nao.ac.jp; Present address: 
National Astronomical Observatory of Japan, Mitaka, Tokyo 181-8588, Japan} 
Paul D.~Dobbie,$^{2}$ Takahiro Nagayama,$^{1}$ Yasushi Nakajima,$^{1}$ 
\newauthor %
Tetsuya Nagata,$^{1}$ Motohide Tamura,$^{3}$ Tadashi Nakajima,$^{3}$ 
Koji Sugitani,$^{4}$
\newauthor %
Hidehiko Nakaya,$^{5}$ Simon T.~Hodgkin,$^{6}$ Andrew J.~Pickles,$^{7}$ 
and Shuji Sato$^{1}$\\
$^{1}$Department of Astrophysics, Nagoya University, Chikusa-ku, 
Nagoya 464-8602, Japan\\
$^{2}$XROA Group, Department of Physics and Astronomy, University of 
Leicester, University Road, Leicester LE2 7RH, UK\\
$^{3}$National Astronomical Observatory of Japan, Mitaka, 
Tokyo 181-8588, Japan\\
$^{4}$Institute of Natural Science, Nagoya City University, Mizuho-ku, 
Nagoya 467-8501, Japan\\
$^{5}$Subaru Telescope, National Astronomical Observatory of Japan, 
Hilo, HI 96720, USA\\
$^{6}$Institute of Astronomy, University of Cambridge, Madingley Road, 
Cambridge CB3 0HA, UK\\
$^{7}$Institute for Astronomy, University of Hawaii, Hilo, HI 96720, USA
}
\begin{document}

\date{Updated on 2003.3.10}

\pagerange{\pageref{firstpage}--\pageref{lastpage}} \pubyear{2003}

\maketitle

\label{firstpage}

\begin{abstract}
We have carried out a brown dwarf search over an area of $14\arcmin \times 23\arcmin$ 
near the central portion of the Pleiades open cluster in five optical and near infrared 
bands ($i', Z, J, H, Ks$) with 10 $\sigma$ detection limits of $i' \sim$22.0, 
$J \sim$20.0 and $Ks \sim$18.5 mag.
The surveyed area has large extinction in excess of $\rm A_V$ = 3 in the Pleiades region.
We detected four new brown dwarf candidates from the colour-colour ($J-K, i'-J$) 
and the colour-magnitude ($J, i'-K$) diagrams. 
We estimated their masses as 0.046 $\rm M_{\sun}$ down to 0.028 $\rm M_{\sun}$. 
The least massive one is estimated to have a mass smaller than Roque 25 \citep{mar98} 
or int-pl-IZ-69 \citep{dob02b}, and maybe the lowest mass object found so far in the 
Pleiades cluster.
\end{abstract}

\begin{keywords}
open clusters and associations: individual (Pleiades) -- 
stars: low-mass, brown dwarfs
\end{keywords}

\section{Introduction}

Many brown dwarf (BD) searches have targeted young open clusters or star forming 
regions rather than general fields, because here BDs are still relatively bright and warm during 
their early phase. 
The Pleiades region is one of the best clusters for this purpose; 
it is a fairly young, nearby, rich and compact cluster. 
It has an age of 125 Myr \citep{sta98a} at a distance of 130 pc \citep{cra76, pin00}. 
Twelve hundred member stars are located within 2.5 degrees of the 
cluster centre \citep{pin98}. 

A number of optical surveys searching for BDs in the Pleiades cluster have been done to 
date and identified a numerous population of bona-fide BDs and BD candidates; 
``CFHT-PL'' objects found by an $RI$ survey \citep{bou98};
``NPL'' objects by an $RIJK$ survey \citep{fes98};
``MHObd'' objects by a $VI$ survey \citep{sta98b};
``IPMBD'' objects by an $RI$ survey \citep{ham99};
``Roque'' objects by an $IZ$ survey \citep{zap99};
``BPL'' objects by an $IZ$ survey \citep{pin00};
``int-pl-IZ'' objects by an $IZ$ survey \citep{dob02b}.
About 20 of these objects are confirmed as Pleiades brown dwarfs on the basis of 
observed lithium abundance, radial velocity and proper motion 
\citep[e.g.,][]{reb95, reb96, bas96, zap97a, sta98a, mar00}. 
However, to date, only one L-type BD candidate \citep[Roque25;][]{mar98} has been 
spectroscopically identified in the Pleiades cluster.
Detection of L-type BDs is difficult at optical wavelengths because 
they should emit most of their bolometric flux in the infrared region.
We have carried out a brown dwarf survey in both the optical and near infrared 
of a small region of the central region of the Pleiades cluster or southern portion of Merope.
Most existing surveys have ignored this region because of its proximity to 
a small molecular cloud which results in large and variable extinction \citep{bre87}.

\section{Observations}

\begin{figure}
	\includegraphics[width=\linewidth]{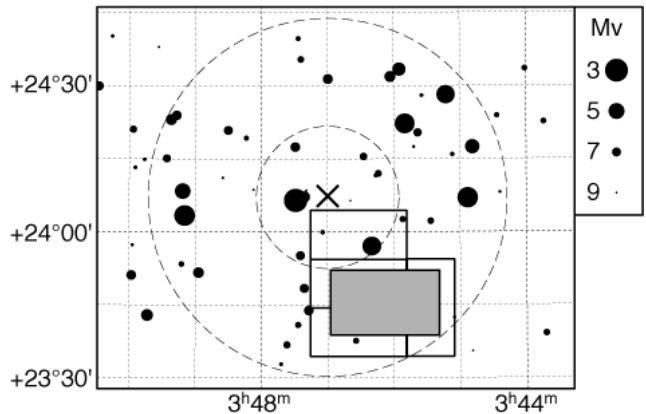}
	\caption{The field locations on the sky.
	The shaded square shows the observed area 
	and the open squares show the region covered only in optical observation.
	The overplotted dashed circles are radii of $0.26\arcdeg$ and $0.64\arcdeg$ 
	from the centre of the cluster, $\rm R.A. 3^h 47^m, Dec.+24\arcdeg 07\arcmin$
	\citep[J2000.0; ][]{pin00}.
	\label{fig:ObsMap}}
\end{figure}

We surveyed an area of $14\arcmin \times 23\arcmin $ centred at 
$\rm R.A. 3^h 46^m 59^s.6, Dec.+23\arcdeg 44\arcmin 59\arcsec (J2000.0)$ 
both in the optical and near infrared bands. 
The observed area is shown as the shaded rectangle in Fig.~\ref{fig:ObsMap}. 
The four open rectangles show the area covered at optical wavelengths.

\subsection{Near Infrared Survey}

Our near infrared survey was made on 2000 October 12, 17 and 18 with 
SIRIUS (Simultaneous InfraRed Imager for Unbiased Survey) on the 
University of Hawaii 2.2 m telescope, atop Mauna Kea. 
SIRIUS employs three 1024 $\times$ 1024 pixel HgCdTe arrays which 
provides $J$ (1.25$\mu$m), $H$ (1.65$\mu$m), and $Ks$ (2.15$\mu$m) 
band images simultaneously \citep{nag99,nag02}.
The field of view in the each band is $4\farcm 9 \times 4\farcm9$ with a pixel scale of 
$0\farcs 28$ at the Cassegrain focus of f/10. 
We mapped an area of $14\arcmin \times 23\arcmin$ with 3 $\times$ 5 tiles of the 
field of view. 
We obtained each of the tiles by dithering 18 frames. 
The exposure time was 60 sec for each frame. 
The limiting magnitudes at 10 $\sigma$ are 20.0, 19.2, and 18.5 at $J$, $H$, and $Ks$, 
respectively.
Typical seeing in the $Ks$ band was $1\farcs 0$.

Dark and dome-flat frames were taken at the beginning or the end of each night. 
We observed standard stars, Nos. 9105, 9107, 9108, 9116 and 9188, in the faint NIR 
standard star catalog of \citet{per98} on the same nights for photometric calibration, 
and the red standard stars, LDN 547, BRI B0021-0214 and BRI B2202-1119, 
in Table 3 of \citet{per98} on 2001 August 31 for determination of the colour 
conversion formula from ours to the CIT system. 
We obtained the following equations as colour-conversion from ours to the CIT system. 
\begin{eqnarray*}
J_{\rm CIT}\!=\!J_{\rm obs}\!-\!(0.013\!\pm\!0.017)\!\!\times\!\!(J\!-\!Ks)_{\rm obs}
\!+\!(0.012\!\pm\!0.030)\\
H_{\rm CIT}\!=\!H_{\rm obs}\!-\!(0.029\!\pm\!0.029)\!\!\times\!\!(J\!-\!H)_{\rm obs}
\!+\!(0.007\!\pm\!0.033)\\
K_{\rm CIT}\!=\!Ks_{\rm obs}\!-\!(0.024\!\pm\!0.020)\!\!\times\!\!(J\!-\!Ks)_{\rm obs}
\!+\!(0.018\!\pm\!0.035)
\end{eqnarray*}

We used pipeline software based on NOAO's IRAF (Imaging Reduction \& 
Analysis Facility)\footnote{
IRAF is distributed by the National Optical Astronomy Observatory, which are 
operated by the Association of Universities for Research in Astronomy, Inc.,
under cooperative agreement with the National Science Foundation.} 
package to reduce the data. 
We have applied standard NIR image reduction procedures, including 
dark current subtraction, sky subtraction and flat fielding. 
Source detection, morphological classification and photometry were performed 
with the APPHOT package in IRAF.

\subsection{Optical Survey}

The $i'$ (0.77 $\mu$m, SDSS system) and $Z$ (0.9 $\mu$m, RGO) band image was 
taken on 2000 October 20 with the Wide Field Camera (WFC) on the Isaac Newton 
Telescope (INT) at La Palma, as part of the INT Wide Angle Survey \citep{mcm01}. 
The WFC accommodates four 2048 $\times$ 4196 pixel EEV CCDs. 
The field of view for each CCD is $11\farcm 4 \times 22\farcm 8$ with a pixel scale of 
$0\farcs 33$ at the prime focus of the INT and the mosaic layout of the four CCDs enables 
us to encompass the whole field of $14\arcmin \times 23 \arcmin$ taken in the near 
infrared bands (Fig.~\ref{fig:ObsMap}). 
We obtained one frame for each filter; the exposure time 
of each frame was 600 s.
The 10 $\sigma$ detection limits are 22.0 and 20.5 at $i'$ and $Z$, respectively, 
where $i'$ and $Z$ are on natural system of WFC.

The observed data were reduced at the Institute of Astronomy, Cambridge, 
using the WFC data reduction pipeline software \citep{irw01}. 
All the frames were bias-subtracted and then corrected for linearity 
before the science fields were flatfielded and defringed. 
The pipeline was also used to generate photometrically and astrometrically 
calibrated object catalogues. 
On the night in question, the $i'$- and $Z$-band photometric zeropoints were measured 
from three standard star fields (observed once each) in the catalogue of 
\citet{lan92} and were found to be stable to about 1 \%. 
The astrometric solution takes account of the radial distortion across the WFC 
field of view, and resulted in residuals of $0\farcs 22$.
\section{Selection of Pleiades BD Candidates}

\subsection{The colour-colour diagram}\label{sec:TCD}

\begin{figure*}
	\includegraphics[width=0.8\linewidth]{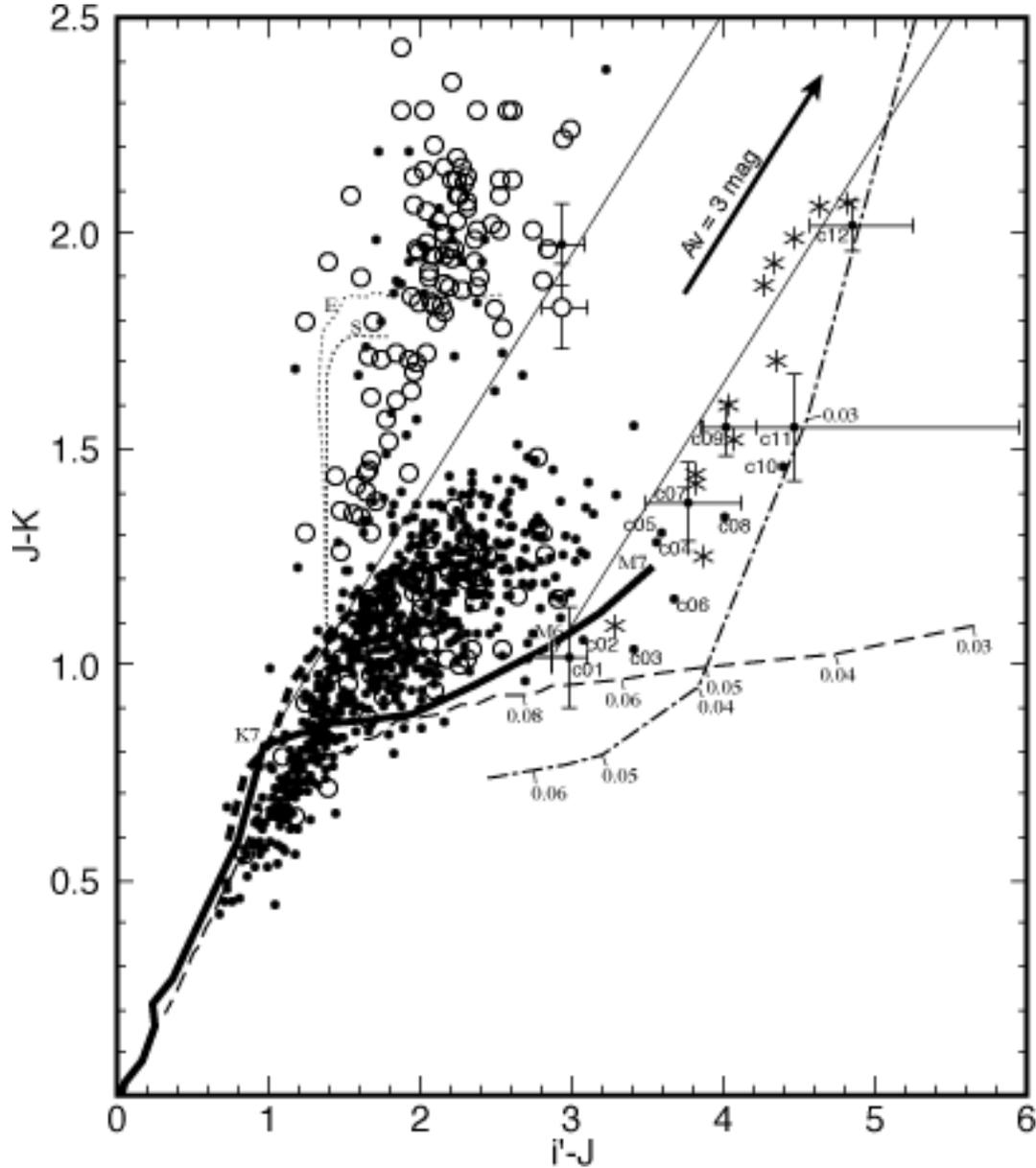}
	\caption{The $J-Ks, i'-J$ colour-colour diagram for all objects detected in all the 5 bands.
	Small dots are point sources and open circles are extended sources.
	Asterisks are the known late-type field dwarfs.
	The thick solid and dashed curves from the bottom left corner represent the loci 
	of main sequence and red giant stars.
	The arrow at the upper right corner shows the reddening vector of $\rm A_V$ = 3 mag.
	The thin dashed line is the $\sim$125 Myr isochrone of the \textsc{NextGen} model 
	\citep{bar98}, and the thin dot-dashed line is the $\sim$120 Myr isochrone of the 
	\textsc{Dusty} model \citep{cha00}. 
	Numbers with tick marks denote masses in solar mass units for each model.
	\label{fig:TCD}}
\end{figure*}

We constructed the colour-colour ($J-K, i'-J$) diagram for all the objects, 
846 sources in total, detected with S/N ratios higher than 10 in the near infrared 
and optical bands, 
differentiating them as extended (149; open circles) and point (697; filled circles) 
sources in Fig.~\ref{fig:TCD}. 
Seven sources detected with S/N ratios higher than 10 in the near infrared but 
lower than 5 in the optical are also plotted with large error bars. 
``Extended objects'' were sorted with a criterion of the FWHMs larger by at least 20 \% 
than the seeing size of the night in the $Ks$ images. 

The thick solid and dashed curves from the bottom-left corner represents the 
loci of main sequence and red giant stars \citep{bes91,bes79}; 
spectral types of K7 dwarf at the cusp at around (1,0.8) and of M7 
at the tip (3.5,1.2) for ($i'-J, J-K$), 
while the arrow at the upper-right corner shows the reddening vector of 
$\rm A_V $ = 3 mag \citep{sch75,rie85}. 
Most of the point sources (filled circles) are concentrated above the locus from 
K7 to M7, along the extinction vector up to 3 mag from the main sequence (reddening band), 
indicating they are K- or M-type dwarfs lying intrinsically on the main sequence. 
On the other hand, most of the extended sources and some point sources spread 
upward of the reddening band. 
About 20 of the extended sources have galaxy-like shapes. 
The K-correction curves from z = 0 - 0.1 of elliptical and 
spiral galaxies (Furusawa, private communication) are plotted as dotted lines.
Thus, objects spread upward of the reddening band including the point sources 
(182 objects in total) are likely to be elliptical or spiral type galaxies reddened with 
K-corrections of z $\sim$ 0 to 0.1 and $\rm A_V $ of 0 to several magnitudes.  
A major problem in this kind of survey is to distinguish stars from 
distant galaxies at faint magnitudes \citep[e.g.][]{fes97}.
The optical and near-infrared colour-colour diagram allows us to easily distinguish them. 
Note that the near infrared colour-colour diagram ($J-H, H-K$) alone cannot distinguish 
low-mass stars from these galaxies, due to merging of them in the diagram 
\citep[e.g.][]{kir99,jar00}.

To highlight the expected location of Pleiades BDs in the colour-colour diagram,
two theoretical isochrones for the Pleiades cluster are also plotted in Fig.~\ref{fig:TCD}; 
the dashed line is the 125 Myr isochrone of the \textsc{NextGen} model \citep{bar98}, 
while the dot-dashed line is the 120 Myr isochrone of the \textsc{Dusty} model 
\citep[; Baraffe, private communication for 120 Myr model]{cha00}. 
The $I_{\rm C}$ magnitudes of these models have been transformed on to our system 
using the following relation, $i' = I_{\rm C} + 0.211(R_{\rm C}-I_{\rm C}) + 
0.011(V-I_{\rm C})^2-0.003(V-I_{\rm C})$.
This transformation is derived via Landolt system ($I_{\rm Lan}$); 
the transformation between $i'$ and $I_{\rm Lan}$ is from an INT web page
\footnote{http://www.ast.cam.ac.uk/\textasciitilde wfcsur/photom.php},
and that between $I_{\rm Lan}$ and $I_{\rm C}$ is from \citet{bes90}.
The \textsc{NextGen} model is based on the nongray dust-free atmosphere \citep{hau99}, 
and describes successfully various observed properties of early-mid M dwarfs, 
but predicts near infrared colours too blue compared to the observed ones for cooler objects. 
The \textsc{Dusty} model includes the effect of grain formation, that is, 
1) the photospheric depletion of dust-forming elements, and 
2) scattering and absorption by dust, explaining successfully the near infrared colours of the 
late M and L types with 1,800 $\lesssim \rm T_{eff} \lesssim 2,200$ K \citep[e.g.][]{jam02}. 
Because of the inadequacy of the \textsc{NextGen} model in the cooler BD region, 
we have adopted the \textsc{Dusty} model in the following discussion.
Using near infrared photometry and lithium tests of low mass cluster members, \citet{zap97b} 
estimated the stellar / substellar boundary to be located at an absolute magnitude of 
($\sim$ 0.075 $\rm M_{\sun}$) for Pleiades ($\sim$120 Myr) to be at $\rm M_I$ = 12.4, 
$\rm M_J$ = 10.0 and $\rm M_K$ = 9.0 mag.
This corresponds to the point ($i'-J, J-K$) = (2.9, 1.0) in our colour-colour diagram, 
highlighted by a cross, at approximately spectral type M6. 
Twelve late-type field dwarfs from \citet{dob02a} were also plotted as asterisks. 
We adopt the redder side of the line running from the cross in parallel 
with the extinction vector as the location of Pleiades BDs in Fig.~\ref{fig:TCD}. 
Note that this location coincides with that of late-M and L-type field dwarfs.
There remain 12 sources on the redder side of the line. 

\subsection{The colour-magnitude diagram}

\begin{figure*}
	\includegraphics[width=0.7\linewidth]{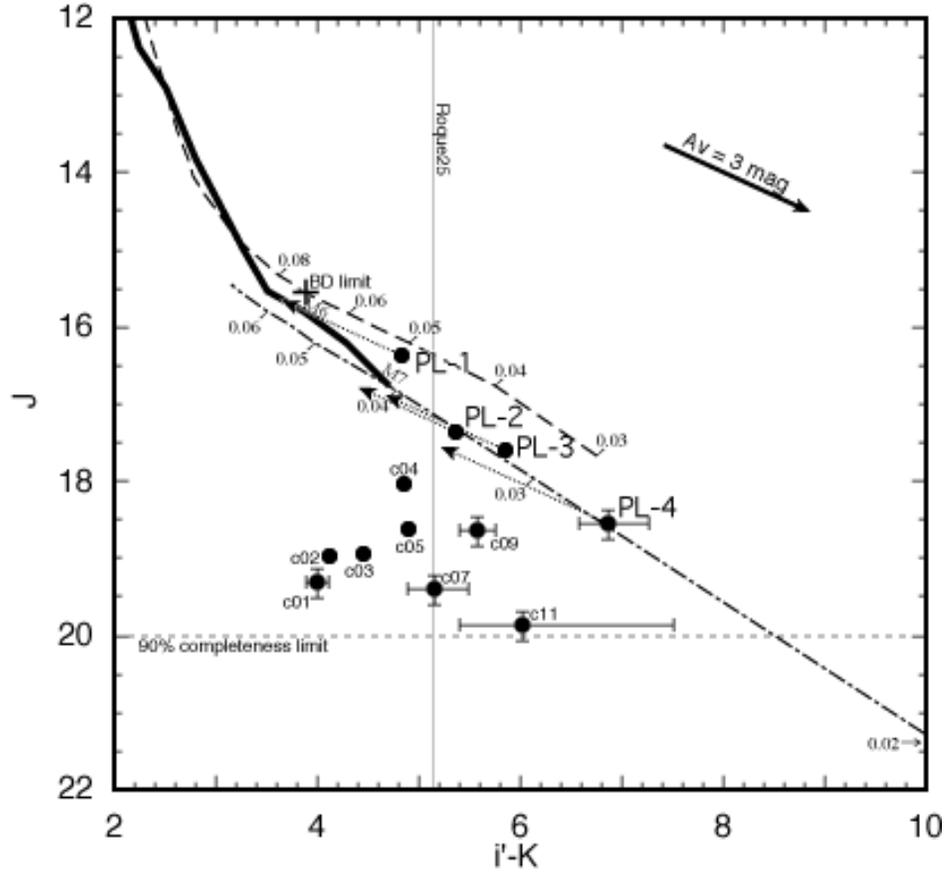}
	\caption{The $J, i'-K$ colour-magnitude diagram.
	The BD candidates are shown with filled circles. 
	Dotted arrow from each candidates shows the estimated maximum 
	reddening vector (see Section~\ref{sec:Av}).
	Other symbols are as in Fig.~\ref{fig:TCD}.
	The $i'-K$ colour of Roque 25 \citep{mar98} is also shown with a perpendicular line 
	due to lack of $J$ magnitude.
	\label{fig:CMD}}
\end{figure*}

To separate Pleiades BDs from field dwarfs, we construct the colour-magnitude 
diagram ($J, i'-K$). 
The 12 red objects selected by Fig.~\ref{fig:TCD} are plotted in Fig.~\ref{fig:CMD}.
The theoretical tracks representative of the low mass stellar and substellar members of 
the Pleiades are also plotted; 
\textsc{NextGen} model \citep{bar98} and \textsc{Dusty} model \citep{cha00}.
We choose the $J$ magnitude as the ordinate of the diagram because $i'$ magnitudes 
of the red objects have larger errors and $K$ magnitudes are less sensitive to the 
mass difference on the theoretical tracks.
The arrow at the upper-right corner indicates the reddening vector of $\rm A_V $ = 3 mag,
which runs nearly in parallel with the tracks.

We adopt the distance modulus used by \citet{dob02b}, $(m-M)_0=5.53$, to estimate the 
ordinate displacement ($J$ magnitude) of the isochrone.
Different age estimates (70 -- 150 Myr) result in a displacement of $\sim$ -0.3 and 
$\sim$ +0.1 mag.
Different distance estimates result in a displacement of $\sim$ -0.2 mag 
(\textit{Hipparcos}) and $\sim$ +0.1 mag (photometry).
The effect of the cluster depth results in a displacement of $\sim \pm$ 0.2 mag.
Additionally, location of unresolved binaries is by 0.75 mag above the single-star 
sequence \citep[e.g.][]{pin00}.
Accounting for all of these uncertainties and allowing for a small degree of error in both 
of the theoretical models and photometry, BDs belonging to the Pleiades cluster should be 
located between 0.3 mag below and 1.0 mag above the relevant single-star isochrone.

\begin{figure}
	\includegraphics[width=\linewidth]{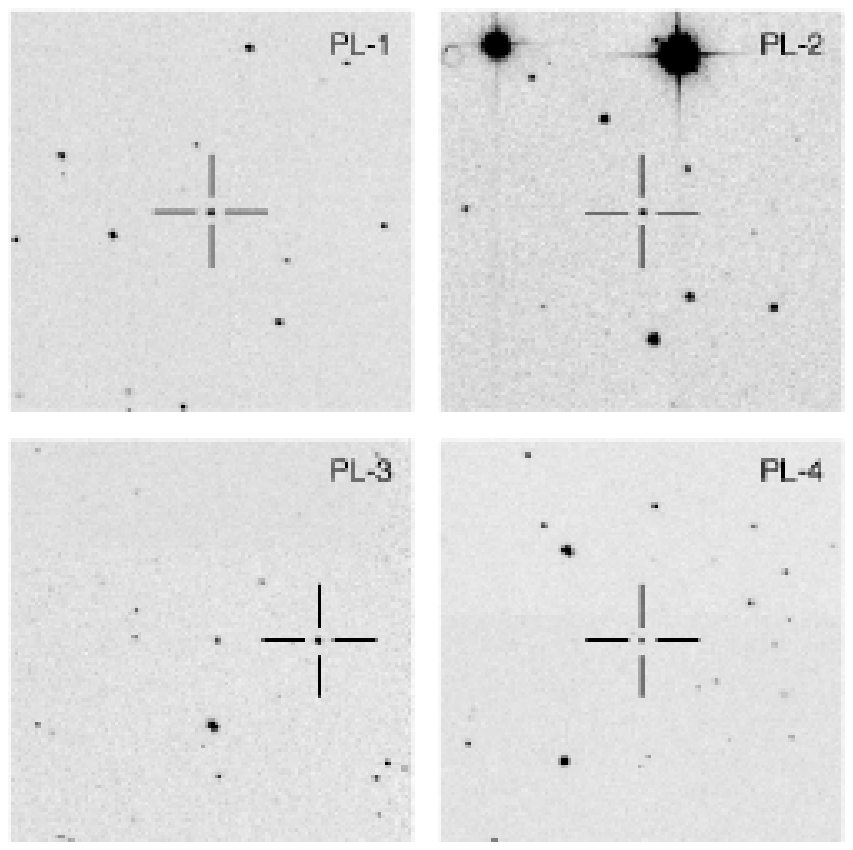}
	\caption{Finding charts for new brown dwarf candidates ($Ks$ band). 
	Each panel is $2\arcmin \times 2\arcmin$, North is up, East is left.
	\label{fig:FindChart}}
\end{figure}

With these criteria, we identified four Pleiades BD candidates from 12 red objects.
The remaining eight objects lie more than 1 mag below the isochrone in the 
colour-magnitude diagram, although they fall in the Pleiades BD domain in the 
colour-colour diagram. 
They could be late-M and L-type dwarfs lying background of Pleiades. 
Finally, we list the four sources (PL-~\# of Fig.~\ref{fig:CMD}) as BD candidates from 
our five-colour survey.
All of them are previously unpublished. 
The positions and magnitudes of four BD candidates are provided in Table~\ref{tab:phot}, 
and that of the remaining eight objects are in Table~\ref{tab:phot2}.
Finding charts of four BD candidates are presented in Fig.~\ref{fig:FindChart}.

\begin{table*}
	\centering
	\begin{minipage}{160mm}
	\caption{The positions and magnitudes of new brown dwarf candidates. 
		\label{tab:phot}}
	\begin{tabular}{ccccccccc}
	\hline
	ID 	& R.A. 	& Dec.		 & $i'$ 				& $Z$ 		
		& $J$ 				& $H$				& $K$ 		& name\\
	\hline
	c06 & 03:45:50.6 & +23:44:37 & 20.02 $\pm$ 0.02	& 18.19 $\pm$ 0.01
		& 16.34 $\pm$ 0.06 	& 15.68 $\pm$ 0.07 	& 15.19 $\pm$ 0.07 & PL-1\\
	c08 & 03:46:34.3 & +23:50:04 & 21.36 $\pm$ 0.06 	& 19.52 $\pm$ 0.03
		& 17.35 $\pm$ 0.06	& 16.63 $\pm$ 0.07	& 16.01 $\pm$ 0.08 & PL-2\\
	c10 & 03:45:11.7 & +23:41:44 & 21.98 $\pm$ 0.09	& 19.88 $\pm$ 0.05
		& 17.59 $\pm$ 0.06	& 16.79 $\pm$ 0.07	& 16.13 $\pm$ 0.08 & PL-3\\
	c12 & 03:45:58.5 &+23:41:54 & $23.40^{\ +\ 0.40}_{\ -\ 0.29}$ 
											& $21.69^{\ +\ 0.29}_{\ -\ 0.23}$
		& 18.54 $\pm$ 0.08 	& 17.42 $\pm$ 0.09	& 16.53 $\pm$ 0.09 & PL-4\\
	\hline
	\end{tabular}
	\end{minipage}
\end{table*}

\begin{table*}
	\centering
	\begin{minipage}{160mm}
	\caption{The positions and magnitudes of other red objects. 
		\label{tab:phot2}}
	\begin{tabular}{cccccccc}
	\hline
	ID 	& R.A. 	& Dec.		 & $i'$ 				& $Z$ 		
		& $J$ 				& $H$				& $K$ \\
	\hline
	c01 & 03:46:24.9 & +23:46:10 & $22.28^{\ +\ 0.12}_{\ -\ 0.11}$ 
											& $20.93^{\ +\ 0.13}_{\ -\ 0.12}$ 
		& 19.30 $\pm$ 0.08 	& 18.90 $\pm$ 0.12	& 18.28 $\pm$ 0.13\\
	c02 & 03:45:30.9 & +23:49:51 & 22.03 $\pm$ 0.1	& 20.45 $\pm$ 0.08 
		& 18.97 $\pm$ 0.07	& 18.32 $\pm$ 0.08	& 17.91 $\pm$ 0.09 \\
	c03 & 03:45:59.7 & +23:40:53 & 22.33 $\pm$ 0.14 	& 20.82 $\pm$ 0.12
		& 18.93 $\pm$ 0.07	& 18.35 $\pm$ 0.08	& 17.89 $\pm$ 0.10\\
	c04 & 03:46:26.6 & +23:39:35 & 21.57 $\pm$ 0.07	& 20.14 $\pm$ 0.06
		& 18.01 $\pm$ 0.06	& 17.22 $\pm$ 0.07 	& 16.72 $\pm$ 0.08\\
	c05 & 03:46:28.0 & +23:42:39 & 22.19 $\pm$ 0.12	& 21.11 $\pm$ 0.15 
		& 18.60 $\pm$ 0.07 	& 17.74 $\pm$ 0.08	& 17.29 $\pm$ 0.09\\
	c07 & 03:46:20.7 & +23:41:56 & $23.18^{\ +\ 0.35}_{\ -\ 0.27}$
											& $22.18^{\ +\ 0.52}_{\ -\ 0.35}$ 
		& 19.14 $\pm$ 0.08	& 18.75 $\pm$ 0.10	& 18.03 $\pm$ 0.11\\
	c09 & 03:45:50.9 & +23:44:57 & $22.65^{\ +\ 0.19}_{\ -\ 0.16}$ 
											& $20.84^{\ +\ 0.13}_{\ -\ 0.12}$ 
		& 18.63 $\pm$ 0.07	& 17.64 $\pm$ 0.08	& 17.08 $\pm$ 0.09\\
	c11 & 03:45:49.2 & +23:41:26 & $24.34^{\ +\ 1.49}_{\ -\ 0.61}$ 
											& $21.93^{\ +\ 0.26}_{\ -\ 0.21}$
		& 19.88 $\pm$ 0.10	& 18.78 $\pm$ 0.13	& 18.33 $\pm$ 0.14\\
	\hline
	\end{tabular}
	\end{minipage}
\end{table*}
\section{Discussion}

\subsection{Estimation of the BD candidate masses} \label{sec:Av}

To estimate masses of BD candidates from their observed magnitudes,
we should determine the level of interstellar extinction towards them.
The distribution of the background field stars in Fig.~\ref{fig:TCD} indicates 
that the interstellar extinction in the surveyed portion of Pleiades ranges from 
0 to 3 mag in $\rm A_V$. 
The foreground extinction toward Pleiades is relatively uniform with the value of 
$\rm A_V \sim$ 0.12~mag, except for a region south of Merope,
which contains a small molecular cloud \citep{cra76}. 
Because the cloud is assigned to be inside the cluster from polarimetry 
and reddening of the member stars and the field stars \citep{bre87}, 
the extinctions estimated from background stars should be regarded 
as upper limit for the BD candidates.

\begin{figure}
	\includegraphics[width=\linewidth]{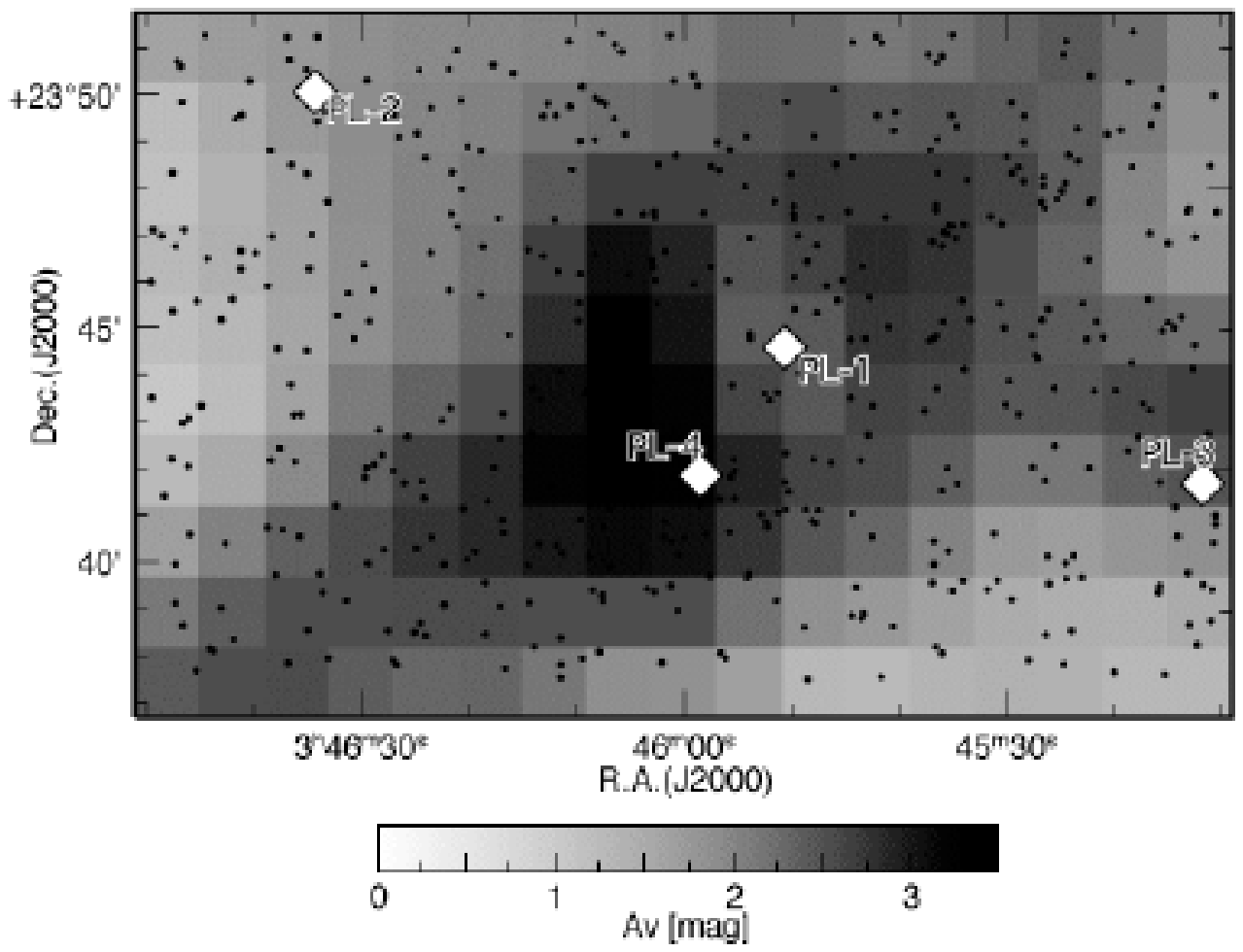}
	\caption{The extinction map constructed from the $J-K$ and $i'-J$ colours 
	of the point sources in the reddening band.
	Small dots are field stars and diamonds are BD candidates.
	Estimation of the extinction is described in Section~\ref{sec:Av}.
	The correspondence between grey scale and $\rm A_V$ is indicated in 
	the scale bar.
	\label{fig:AvMap}}
\end{figure}

To estimate the maximum interstellar extinction for the individual BD candidates, 
we constructed an extinction map in the surveyed area from all the field stars 
(small dots between the two straight lines in Fig.~\ref{fig:TCD}), assuming them 
to be K- or M-type main sequence stars (the thick line in Fig.~\ref{fig:TCD}). 
Visual extinction, $\rm A_V$, toward each star was calculated from its $i'-J$ and 
$J-K$ colours with the reddening law of \citet{sch75} and \citet{rie85}, 
i.e. ${\rm E}(J-K)/{\rm E}(i'-J) = 0.56, {\rm A_V} = 6.10 \times [(J-K)_{\rm obs} 
- (J-K)_{\rm intrinsic}]$. 
We meshed the area with 17 $\times$ 10 grids of $1\arcmin$ square and 
adopted the maximum $\rm A_V$ within the grid, and averaged them over 
each mesh with Gaussian weights. 
The final $\rm A_V$ map is shown in Fig.~\ref{fig:AvMap} and the values 
toward the BD candidates are listed in Table~\ref{tab:mass}.

The estimated maximum extinctions for individual candidates are shown in 
Fig.~\ref{fig:CMD} with dotted arrows.
If extinction is neglected, PL-2 and -3 are estimated to have M $\sim$ 0.035 
$\rm M_{\sun}$, similar to Roque 25 \citep{mar98} and int-pl-IZ-69, -81, -84 
\citep{dob02b}, the coolest candidate members so far detected. 
On the other hand, assuming the maximum likely level of extinction,
their masses are estimated to be M $\sim$ 0.04 $\rm M_{\sun}$.
Similarly we estimate the mass of the faintest candidate to lie in the range of 
0.033 -- 0.028 $\rm M_{\sun}$. 
Therefore, PL-4 maybe the lowest mass member of the Pleiades identified to date.
The estimated masses from $J$ magnitudes for the case of no extinction and 
that of maximum possible extinction toward individual candidates are shown in 
Table~\ref{tab:mass}. 
Note that uncertainties in the age and the distance of the cluster and the effect of the 
cluster depth contribute an additional error of 10 -- 20 \% in the mass estimation; e.g. the 
mass of PL-4 for the case of no extinction is $\rm 0.028^{+0.002}_{-0.001} \ M_{\sun}$.
We note that our mass estimates would increase slightly if we were to take into account the 
``missing M dwarf gap'' \citep{dob02c}. 

\begin{table}
	\caption{Mass estimates for brown dwarf candidates based on 120 Myr 
		DUSTY model. \label{tab:mass}}
	\begin{tabular}{cccc}
	\hline
			&		&Mass($\rm M_{\sun}$)	& \\
	Name 	& MAX $\rm A_V$ 
					& $\rm A_V$=0 -- MAX 	& SpT\\
	\hline
	PL-1 	& 2.5 	& 0.046 -- 0.063 		& M6 -- M8\\
	PL-2 	& 2.0 	& 0.035 -- 0.040 		& M8 -- L1\\
	PL-3 	& 2.5 	& 0.033 -- 0.039 		& M8 -- L1\\
	PL-4 	& 3.5 	& 0.028 -- 0.033 		& L1 -- L5\\
	\hline
	\end{tabular}
\end{table}
\subsection{Broad-band energy distributions}

\begin{figure}
	\includegraphics[width=\linewidth]{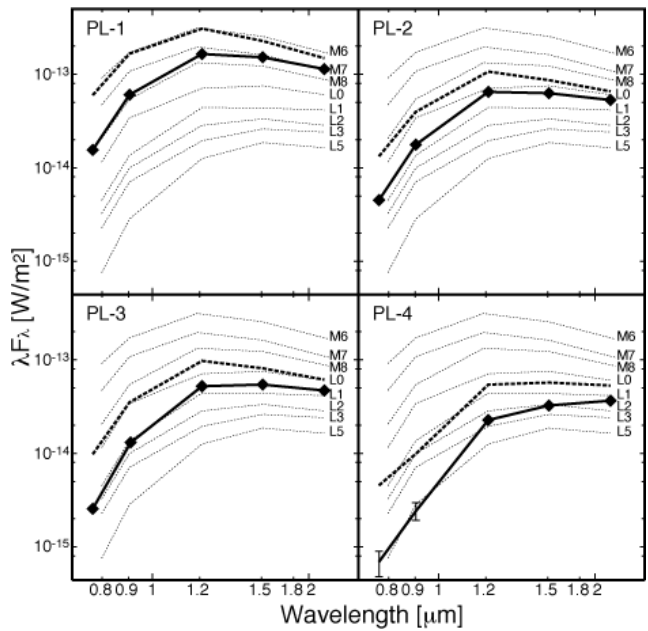}
	\caption{The broad-band energy distributions for four brown dwarf candidates.
	Solid and dashed lines represent brown dwarf candidates with no extinction 
	and with possible maximum extinction, respectively.
	M6 - M8 and L0 - L5 field dwarfs are overplotted with dotted lines.
	\label{fig:SED}}
\end{figure}

In Fig.~\ref{fig:SED} we compare the four SEDs with those of cool field dwarfs,
determined in previous surveys (M6 -- M8; Kirkpatrick \& McCarthy 1994,
 L0 -- L5; Kirkpatrick et al. 1999,2000).
Solid and dashed lines represent SEDs of the four BD candidates for the cases of 
no extinction and maximum extinction discussed in \S~\ref{sec:Av}, respectively. 
We have assumed the distance modulus for Pleiades of $\rm (m-M)_0$ = 5.53 
\citep{cra76, pin00} to convert the observed magnitudes to absolute magnitudes 
for the cluster member.
The zero-magnitude fluxes are taken from \citet{bes98}.

All the candidates show a steep decline towards the shorter wavelengths and a broad 
hump between 1 and 2 $\mu$m.
Within the uncertainties arising from the correction for interstellar extinction, the 
four SEDs of the candidate BDs are well matched by the energy distributions of the 
late-M and L-type field dwarfs: PL-1 is consistent with a mid to late-M spectral type, 
PL-2 and PL-3 with late-M to early L spectral type and PL-4, an early-mid L spectral type. 
We summarize these results in Table~\ref{tab:mass}." 
\subsection{Contamination of field stars}

As discussed in section~\ref{sec:TCD}, the optical and near-infrared colour-colour 
diagram allows us to separate distant galaxies and giants from BD candidates.
However, field dwarfs are still possible sources of contamination.

The \textsc{Dusty} model indicates that Pleiades BDs are overluminous with respect to 
field stars of similar spectral type by $\sim$ 1 mag.
The criterion of selecting objects between 0.3 mag below and 1.0 mag above the 
isochrone means that we are sensitive to single and binary dwarfs at distance of 
71 -- 130 and 96 -- 184 pc, respectively. 
This corresponds to space volumes of 15.5 and 45.1 $\rm pc^3$ for single and 
binary field stars, respectively.
Assuming a space density of M8 -- L4.5 dwarfs of 0.0066 $\rm pc^{-3}$ 
\citep{giz00} and a binary fraction of 50 \% \citep[e.g.][]{ste95}, 
we estimate the contamination of $\sim$ 0.2 field star in our BD candidates.
Thus, the level of the contamination of the BD candidates by non-members is 
negligibly low.

\subsection{The cluster luminosity function and mass function}\label{sec:MF}

\begin{table*}
	\centering
	\begin{minipage}{140mm}
	\caption{$N_{\rm total}$ per unit mass (in the case of $\rm A_V$ = 1.5 mag).
		\label{tab:MF}}
	\begin{tabular}{ccccccc}
	\hline
	J magnitude bin	& Mass bin ($\rm M_{\sun}$) 	& Mid-mass ($\rm M_{\sun}$)
	& $N_{\rm obs}$ &$N_{\rm annul}$ &$N_{\rm total}$ &$N_{\rm total}$ per unit mass\\
	\hline
	15.1 -- 17.1 	& 0.095 -- 0.037 			& 0.066 	
	& 2 			& 26 	& 330		& 5,700 $\pm$ 4,000\\
	17.1 -- 19.6 	& 0.037 -- 0.025 			& 0.031 	
	& 2			& 26 	& 330 		& 28,000 $\pm$ 19,000\\
	\hline
	\end{tabular}
	\end{minipage}
\end{table*}

We have derived the cluster luminosity function for $J$ magnitudes between 15.5 to 
20.0 which correspond to the BD limit of \citet{zap97b} and the 90 \% completeness 
limit of our survey (almost the same as 10 $\sigma$ detection limit), respectively.
In our calculation we assume that the interstellar extinction toward Pleiades in our survey 
area is spatially uniform with the value of 1.5 mag at $V$ or 0.4 mag at $J$, 
we subdivided this magnitude range into 2-mag bins, given in Table~\ref{tab:MF}.

Using the \textsc{Dusty} model, we have calculated the masses corresponding to 
each of our luminosity bins.
We estimated the total number, $N_{\rm annul}$, of member of the whole cluster 
from the number of our sample.
First we calculate the number, $N_{\rm annul}$, of members inside an annulus. 
Our survey area covers 7.7 \% of the area of the annulus of an inner and an outer 
radius of $r_i = 0.26 \arcdeg$ and $r_o = 0.64 \arcdeg$, respectively 
(Fig~\ref{fig:ObsMap}).
$N_{\rm annul}$ is calculated simply from the area ratio.
Next we calculate $N_{\rm total}$.
\citet{pin98} and \citet{rab98} show that the surface density distribution of the 
Pleiades members can be well fitted by a King distribution \citep{kin62} 
whose core radius increases as the stellar mass decreases.
The following equation, which is obtained by integrating the function of King distribution, 
provides the number of cluster members inside the circle of a radius $r$, 
\begin{equation}
n(r) = \pi r^2_ck\left[\ln{(1+x)} - 4\frac{\sqrt{1+x}-1}{\sqrt{1+x_t}} + 
\frac{x}{1+x_t}\right], 
\label{eq:King}
\end{equation}
where $x = (r/r_c)^2$, $x_t = (r_t/r_c)^2$, $k$ is the surface density 
of the cluster, $r$ the radius from the cluster centre, $r_c$ the core radius at 
which the surface density of members falls to half its central value, and $r_t$ 
the tidal radius at which the external gravitational potential equals the cluster potential. 
The total number of cluster members, $N_{\rm total}$, can be determined independently 
of $k$ by taking the ratio $n(r_t) / [n(r_o) - n(r_i)]$, where $n(r_o) - n(r_i)$ is equal to 
$N_{\rm annul}$ and $n(r_t)$ is equal to $N_{\rm total}$.
We adopt a value of $r_c = 2.2 \arcdeg$ from \citet{jam02} for low-mass 
members of Pleiades, and a value of $r_t = 5.54 \arcdeg$ from \citet{pin98}. 
\begin{equation}
\frac{N_{\rm total}}{N_{\rm annul}} = \frac{n(r_t)}{[n(r_o) - n(r_i)]} = 12.78.
\label{eq:ratio}
\end{equation}

\begin{figure*}
	\includegraphics[width=0.6\linewidth]{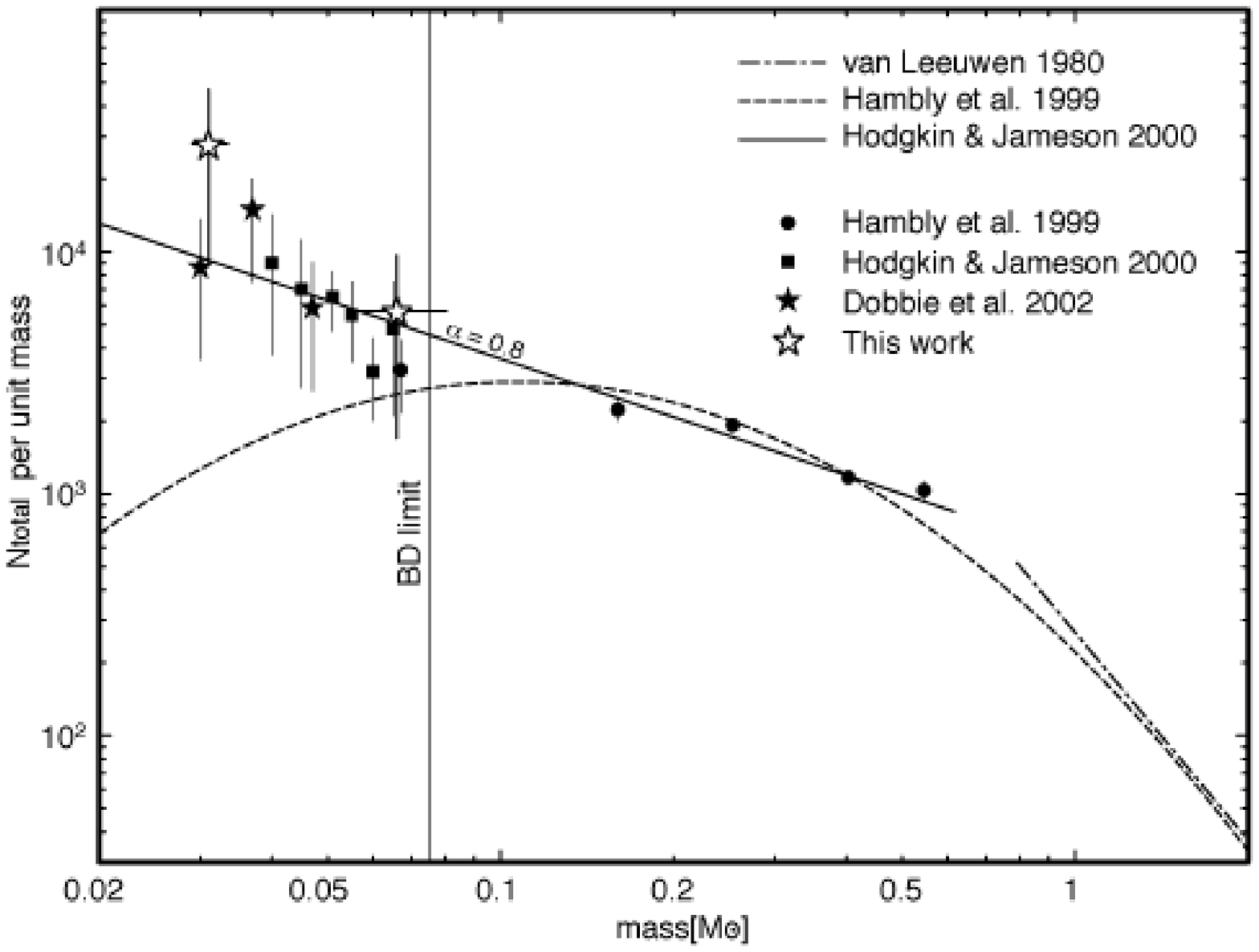}
	\caption{The mass function in the Pleiades cluster. 
	Open stars represent this work.
	Filled stars, squares, and circles represent the results of \citet{dob02b}, 
	\citet{hod00}, and \citet{ham99}, respectively, together with the model fits by 
	\citet{hod00} (solid line), \citet{ham99} (dashed line), \citet{van80} 
	(dot-dashed line).
	\label{fig:MF}}
\end{figure*}

Finally, we calculate the values of substellar mass function of the cluster, 
$N_{\rm total}$ per unit mass.
The values are given in Table~\ref{tab:MF} and are plotted in Fig.~\ref{fig:MF} 
with several previous estimates \citep{dob02b, hod00, ham99}.
The shape of mass function of the Pleiades across and below the stellar / substellar boundary 
has been reported to be well represented by a power-law with an index $\alpha$ lying 
between 0.4 -- 1.0, dependent on investigator \citep[e.g.][]{jam02, bou98, hod00, zap97a}.
Our new points seem to support the larger values of $\alpha$.
However, as the present survey covers only a small fraction of the cluster (i.e., $< 0.5$ \%), 
this result must be treated with a measure of caution.
Nevertheless, we note that \citet{bej01} have recently determined the substellar mass 
function of the much younger $\sigma$ Orionis cluster to be well matched by a power-law 
with index $\alpha$ = 0.8.

\section{Conclusions}

We have undertaken an optical ($i',Z$) and near infrared ($J,H,Ks$) survey of a 
$14\arcmin \times 23\arcmin$ area near the centre of the Pleiades cluster to search 
for brown dwarf members. 
We have unearthed 4 BD candidates (PL-1, -2, -3, and -4), out of 853 sources. 
PL-1 is likely an M-type BD, PL-2 and -3 are likely late-M or early-L type BDs, 
while PL-4 is probably an L-type BD. 
The estimated mass of our faintest candidate, PL-4, is in the range of 0.028 -- 0.033 
$\rm M_{\sun}$, lower than that of Roque 25 \citep{mar98} and can be lower than that 
of int-pl-IZ-69 \citep{dob02b}, possibly making it lowest mass Pleiad identified to date.

\section*{Acknowledgments}

We thank the staffs of the UH 2.2 m telescope and the Subaru telescope 
for supporting the observation of SIRIUS. 
This work is financially supported by \textsc{Sumitomo} foundation, 
Grant-in-Aid for Scientific Researches on Priority Areas (A),
and International Scientific Research, the Ministry of Education, Culture, 
Sports, Science, and Technology.
The Isaac Newton Telescope is operated by the Isaac Newton Group in the Spanish 
Observatorio del Roque de los Muchachos of the Instituto de Astrofisica de Canarias.
Nagashima, Nagayama, and Y.~Nakajima are financially supported by the 
Japan Society for the Promotion of Science.
PDD is a PPARC postdoctral research associate.
We also thank an anonymous referee for his/her useful comments and advice.


\bsp
\label{lastpage}

\end{document}